# Attenuation study for Tibet Water Cherenkov Muon detector array-A

GOU Quanbu[1,*]  GUO Yiqing[1]  LIU Cheng[1]  QIAN Xiangli[1,2]  HOU Zhengtao[1,3]

[1]*Key Laboratory of Particle Astrophysics, Institute of High Energy Physics, Chinese Academy of Sciences, Beijing 100049, China*

[2]*Department of Physics, Shandong University, Jinan 250100, China*

[3]*Department of Physics, Hebei Normal University, Shijiazhuang 050016, China*

**Abstract** This work aims at online calibrating the signal attenuation of the long cable used in Tibet Water Cherenkov Muon detector array-A (Tibet MD-A) under the 37 000 m$^2$ Tibet air shower array. Based on the waterproof connection of the PMT (R3600_06) with the signal cable, and the characters of the high voltage divider, terminal reflection method is explored and used for measuring the signal attenuation with a practical way to eliminate the contribution of the pulse baseline (pedestal). Comparison measurement between it and QDC (Charge-to-Digital Conversion) data taking method are carried out by using open-ended cables, confirming that terminal reflection method is a fast and convenient method, and suitable to online calibrate the signal attenuation for Tibet MD-A. At 26°C, the measured attenuation coefficient with the 250-m cable, permanently connected to the PMT, is ~13.9%. Also, the cable frequency response is measured by using the sinusoidal signals, which could be used to study the time dispersion of the real signal produced with Tibet MD-A detector via Fourier analysis.

## 1 Introduction

Since 1990, the Tibet air shower array experiment[1] has been operating successfully at Yangbajing (YBJ), Tibet, China, located at 90.522_E, 30.102_N, 4 300 m above sea level with an atmospheric depth of 606 g/cm$^2$. In very high energy γ-ray astronomy, cosmic-ray event is a dominant background for the ground-based experiment. In order to discriminate γ-rays from background cosmic-rays, a large underground water-Cherenkov-type Muon-detector array (Tibet MD [2]) under the 37 000 m$^2$ Tibet air shower array is under construction. In this paper, the charge attenuation and influence on the pulse rise time of the transmission line are studied, which are important for the proper working of the data acquisition (DAQ) system.

## 2 Tibet MD-A and the cable attenuation

As one of the 12 Tibet MD detectors, Tibet MD-A has 16 detector units shown schematically in Fig.1 upper panel. Each unit (down panel in Fig.1) contains a 7.2 m × 7.2 m × 1.9 m water tank viewed with a 508-mm PMT (Photomultiplier R3600_06). Each PMT is connected to a DAQ system using a 250-m coaxial cable. The cable consists of a 70-m one-end-loaded cable (RG-58C/U type) with a water-proof seal connection with the PMT, and a 180-m open-ended cable (RG-58/UR type) connected to the DAQ. Both cables belong to RG-58 family with characteristic impedance of 50 Ω, and are interconnected by a BNC (Bayonet Nut Connector) adapter located outside the water tank.

The attenuation measurement is schematically shown in Fig.2. By two reference cables (A1B1 and C1D1), a pulse generator as the signal source is connected to E1, the end of the cable (F1E1) opposite to the PMT, and a signal is sent toward the PMT terminated on 10 kΩ. The signal is almost completely reflected (with a reflection coefficient of up to 99%), and both the direct and reflected signals are recorded on the oscilloscope.

---

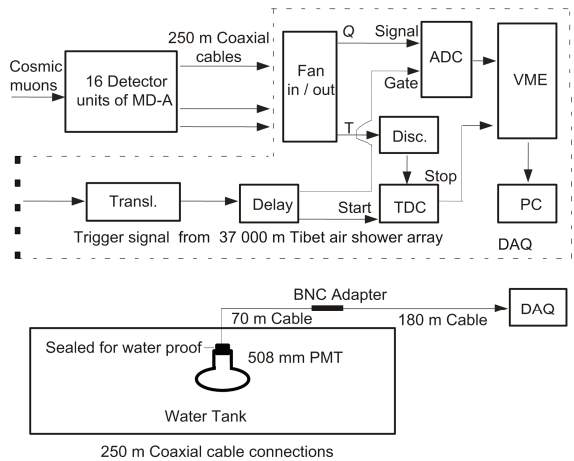

**Fig.1** Experimental setup of MD-A.

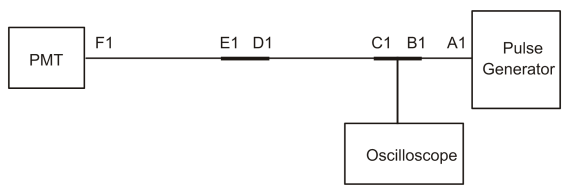

**Fig.2** Diagram of the terminal reflection measurement. A1B1 is a 0.7-m cable to the generator, C1D1 is reference cable, and E1F1 is PMT cable. The Tee adapter connects A1B1, C1D1 and the oscilloscope. The BNC adapter connects C1D1 and E1F1.

The pulse generator has two parallel-output load terminals. One is A1F1 transmission line associated with the PMT terminated on 10 kΩ, and another is the 1-MΩ oscilloscope, from which the source output signal is simultaneously reflected into C1F1 transmission line at the Tee adapter, without affecting the signals.

The generator sent triangular pulses of about 10 ns leading edge, 45 ns trailing edge and 30 ns FWHM, thus roughly reproducing the PMT output signal. The measurement was performed at 26°C.

The cable frequency response for sinusoidal signals was studied by setting the pulse generator in sinusoidal mode. The attenuation coefficient (174 dB/km@100 MHz) was provided by the manufacturer, and used to check the measurement results. As shown in Fig.3, the A2B2 cable connects the pulse generator, and the C2D2 cable (180 m) is to be tested. The sinusoidal input pulse with 5-V amplitude comes from the pulse generator.

The attenuation coefficients in Fig.4 were measured by comparing the signal amplitudes at the input (C2) and the output (D2) of the 180-m cable.

Below a few tens of kHz, the cable frequency response is completely flat. In this region, the measured attenuation is ~9%, which corresponds to an attenuation of ~41% /km at the normalized unit length. This means that at low frequency the attenuation coefficient depends on the cable length instead of frequency.

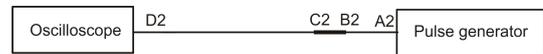

**Fig.3** Diagram of the direct measurement. A BNC adapter connects the A2B2 cable (0.7 m) and test cable (C2D2).

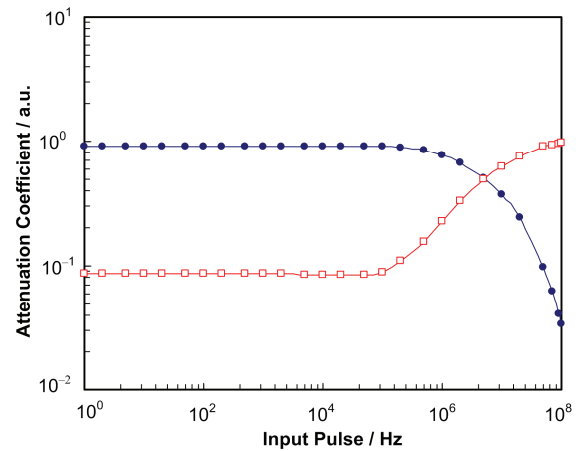

**Fig.4** Attenuation coefficients as a function of pulse frequency.

Above $1\times10^5$ Hz, the attenuation increases quickly with frequency, and reaches the cutoff frequency at about $2\times10^7$ Hz, corresponding to a 17.5-ns rise time, which is roughly consistent with the measured 22-ns rise time of the PMT signal considering that the PMT has the 10-ns nominal rise time .

At $1\times10^8$ Hz the cable attenuation coefficient of the cable is up to 96.6%, and roughly in consistent with 174dB/km@100 MHz (corresponding to 97.3%/ 180 m@100 MHz) provided by the commercial manufacturer.

Also, the measured cable frequency response can be used to study the time dispersion of the real signal produced by Tibet MD-A detector using Fourier analysis.

## 3 Comparison measurement of attenuation coefficient

It is impossible to use the cable end connects the PMT as the source end. The reflection technique at the cable's open-end is used for studying its attenuation.

## 3.1 Terminal reflection method

The reflection coefficient ($\lambda_{rc}$) is given by Eq.(1),

$$\lambda_{rc}=(R-R_{ci})/(R+R_{ci}) \quad (1)$$

where, $R = 10$ kΩ is the terminal load resistance, and $R_{ci} = 50$ Ω is the characteristic impedance of the cable. Considering the PMT output end as the input end, the $R$ is located at the HV-divide board of PMT, and the reflection coefficient approaches to 100%, the long cable meets the requirements of No-Lose terminal reflection.

According to Ref.[3,4], compared with the input pulse, the leading and trailing edges, and the pulse baseline of the reflected pulse are deteriorated. The delay of the reflected pulse depends on the cable length, and is about 2450 ns for 250-m cable. The observed pattern is the sum of the input and the late reflected pulses. To get a reasonable charge of the late reflected pulse, the contribution of the pulse baseline (pedestal) has to be subtracted.

In Fig.2, provided that $Q_{F1C1}$ is the pulse charge after A1F1 and F1C1 cables, and $Q_{D1C1}$ is the pulse charge after A1D1 and D1C1 cables (disconnecting E1F1 cable and PMT at E1). There are reflections at F1 and D1 respectively in these two cases. So the attenuation coefficient $a_{E1F1}$ of E1F1 cable is given by Eq.(2),

$$\alpha_{E1F1}=1-(Q_{F1C1}/Q_{D1C1})^{0.5} \quad (2)$$

where

$$Q = \int_{t1}^{t2} I(t)dt = \int_{t1}^{t2} \frac{U(t)}{R} Rdt$$

At $R=50$ Ω, Eq.(2) can be expressed as

$$a_{E1F1} = 1 - \sqrt{\int_{t1}^{t2} U_{F1C1}(t)dt / \int_{t1}^{t2} U_{D1C1}(t)dt} \quad (3)$$

Then

$$\Delta S = \int_{t1}^{t2} U(t)dt \quad (4)$$

where, ΔS is the pulse area of the observed pattern. The attenuation coefficient of E1F1 cable is given by,

$$\alpha_{E1F1}=1-(\Delta S_{F1C1}/\Delta S_{D1C1})^{0.5} \quad (5)$$

Fig.5 shows the $\Delta S$ measurement. $S_{ub}$ is the area of the baseline unit between two downward arrows, and $S_{refl}$ is the reflected pulse area between two upward arrows. By assuming a time interval of $6S_{ub}$ at $S_{refl}$, the reasonable reflected pulse area ($\Delta S_{refl}$) can be

$$\Delta S_{refl}= S_{refl} - 6S_{ub} \quad (6)$$

## 3.2 Comparison measurement

QDC (Charge-to-Digital Conversion) method can be used to measure the attenuation coefficient for the 180-m open-ended cable, but not for the 70-m one-end-loaded cable. Also, it is important to compare the attenuation coefficient using terminal reflection method with that using QDC facility. The attenuation coefficient of the 180-m open-ended cable was investigated by QDC data taking and terminal reflection methods. The latter is realistic and practical to the conditions at the YBJ site.

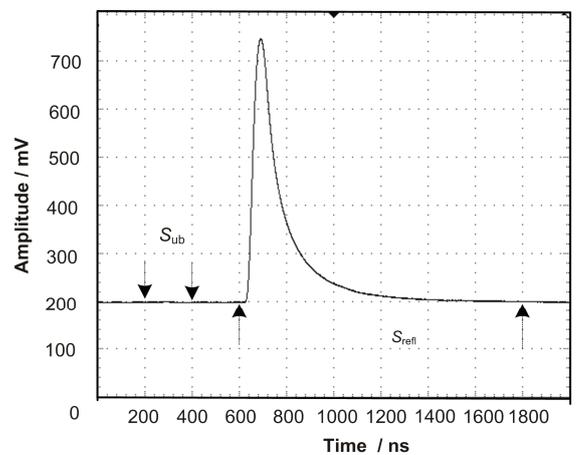

**Fig.5** Measurement of reflected pulse area.

### 3.2.1 Measurement by QDC method

Fig.6 shows the diagram of the QDC method. The square output pulses from the generator to drive LED light source were of 3.9-V amplitude, 700-ps leading and trailing edges, 30-ns pulse width, and 90-Hz repetition.

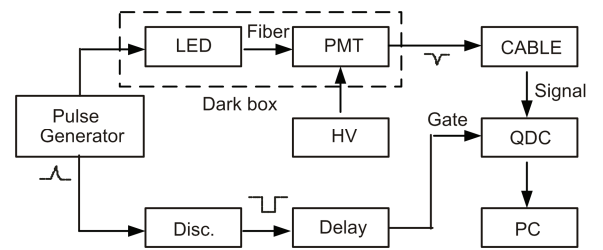

**Fig.6** Diagram of the QDC data taking method.

In order to test the QDC method, the CABLE in Fig.6 was replaced with a 20-dB attenuator. The $Q_{QDC\_20dB}$ (with 20-dB attenuator) was 15.7 pC, and $Q_{QDC\_norm}$ (without 20-dB attenuator) was 152.9 pC. Then, the attenuation coefficient with the 20-dB attenuator ($a_{QDC\_20dB}$) is $a_{QDC\_20dB} = 1 - Q_{QDC\_20dB}/Q_{QDC\_norm} = 89.7\%$. It corresponds to 19.7 dB, and is roughly in agreement with the nominal value of 20 dB. Thus, the attenuation coefficient of the open-ended cable can be measured by the setup in Fig.6.

Again, the CABLE in Fig.6 was replaced by the 180-m open-ended cable. With $Q_{QDC\_180m} = 137.6$ pC, and $Q_{QDC\_norm} = 152.9$ pC, and the attenuation coefficient is $a_{QDC\_180m} = 1 - Q_{QDC\_180m}/Q_{QDC\_norm} = 10\%$.

So, the QDC method is reliable, and can be used to check other measurement methods.

### 3.2.2 Measurement by terminal reflection method

In Fig.2, PMT and the E1F1 cable was replaced by the 180-m open-ended test cable, and the C1D1 was the 180-m reference cable. The waveforms of the pulse generator and the reflected signal were recorded on the oscilloscope.

The triangular output pulses from the generator were of 5-V amplitude, 10-ns leading edge, 45-ns trailing edge, 30-ns pulse width, and 90-Hz repetition. The pulse shape selected by the results in Section 2 is similar to the output signal of the MD-A detector.

Using the terminal reflection method and Eq. (6), we had $\Delta S_{refl\_180m} = 77.62$ nVs, and $\Delta S_{refl\_360m} = 63.28$ nVs. Using Eq. (5), we had the attenuation coefficient $a_{refl\_180m} = 9.7\%$. This is roughly in agreement with $a_{QDC\_180m}$ obtained using QDC data taking method.

### 3.3 Attenuation coefficient with total cable length

According to above results, the attenuation coefficient with the 250 m one-end-loaded cable in Tibet MD-A can be measured by using the terminal reflection method, and the experimental setup schematically shown in Fig.2. For each detector unit in Fig.1, the DAQ system was replaced by the devices after PMT and F1E1 cable. The A1B1 is just used to connect the pulse generator; the C1D1 corresponds to 180-m reference cable; the E1F1 is 250 m (70 m +180 m) one-end-loaded cable under test. The input pulse of the circuits from the pulse generator is the same as in Section 3.1.

For one Tibet MD-A detector unit, using the terminal reflection method and Eq. (6), we had $\Delta S_{refl\_180m} = 78.26$ nVs, and $\Delta S_{refl\_430m} = 58.06$ nVs.

Using Eq.(5), we had the attenuation coefficient $a_{refl\_250m} = 13.9\%$.

Therefore, at 26°C the attenuation coefficient with the 250 m long cable is corrected as 13.9%, corresponding to 0.52 dB/100 m.

## 4 Conclusions

The cable frequency response is measured with the sinusoidal signals. It is completely flat below a few tens of kHz, indicating that its attenuation coefficient at low frequency mainly depends on the cable length but not the frequency. The attenuation above 100 kHz increases quickly with frequency, and reaches the cutoff frequency about 20 MHz, corresponding to a 17.5-ns rise time.

Furthermore, the measured frequency response can be used to do Fourier analysis for the real signal produced with Tibet MD-A detector in the near future. Considering the commercial 508 mm PMT (Photomultiplier, R3600_06) has a waterproof connection with the signal cable and the characters of the high voltage divider of the PMT, we developed terminal reflection method to study the attenuation of the long cable; moreover, the contribution of the pulse baseline (pedestal) is successfully eliminated. The attenuation coefficient of the 180-m open-ended cable obtained with terminal reflection method ($a_{refl\_180m}$, 9.7%) is roughly in agreement with that obtained with QDC data taking method ($a_{qdc\_180m}$, 10%). It confirms that terminal reflection method is reliable and convenient, and can be used to online calibrate the signal attenuation for Tibet MD-A detector. At 26°C, the attenuation coefficient ($a_{refl\_250m}$, 13.9%) of the 250 m long cable connected to the PMT of the Tibet MD-A detector is obtained with terminal reflection method.

In addition to this, we are going to study the signal attenuation caused by the electronics in the DAQ system presented in Fig.1.

## Acknowledgements

The authors would like to sincerely thank Prof. Hu Hongbo, Prof. Shen Changquan, Prof. Zhang Yi and Prof. Wang Hongwei for their strong supports or helpful discussions. The collaborative experiment of the Tibet Air Shower Arrays has been performed under the auspices of the Ministry of Science and Technology of China and the Ministry of Foreign Affairs of Japan. This work is supported by the National Natural Science Foundation of China (No.111 072552430) and the Key Laboratory of Particle Astrophysics, Institute of High Energy Physics, Chinese Academy of Sciences.

## References

1 Amenomori M, Bi X J, Chen D, *et al.* APJ, 2009, **711:** 119–124.

2 Amenomori M, Bi X J, Chen D, *et al.* AIP Conf. Proc. 2009, **1085:** 723–726.

3 http://www.globalspec.com/reference/14847/160210/Chapter-4-Electromagnetic-Compatibility-and-Medical-Devices-Puls-Reflection-and-Termination-Techniques

4 Stuchly M A, Stuchly S S. IEEE Trans Instrum Meas, 1980, **IM-29:** 176–183.